\documentclass[english]{article}
\usepackage[T1]{fontenc}
\usepackage[latin9]{inputenc}
\usepackage[toc]{appendix}
\usepackage{amsmath}
\usepackage{graphicx}
\usepackage{babel}
\usepackage{subcaption}
\usepackage{ifthen}
\usepackage{hyperref}

\newcommand*{\email}[1]{%
  \normalsize\href{mailto:#1}{#1}\par
}

\newcommand*{\norm}[1]{%
  \left|\left|#1\right|\right|
}

\begin{document}
\bibliographystyle{plain}
\title{Equatorial Platform Based on Planer Linkage}
\author{Daniel J Matthews \\ \email{matth036@umn.edu} }

\date{\today}
\maketitle

\section*{Area of Technology}
Visual and photograpic astronomy. Telescope mounts. Solar tracking
for energy collection.

\section*{What is an Equatorial Platform?}

An equatorial platform is a mechanism upon which a telescope on a
non-tracking mount may be placed and the motion of the mechanism will
result in tracking. Here tracking means the object under observation
remains centered in the field of view of the telescope. An equatorial
platform can only track for a limited time period before it must be reset.

Appendix D of reference \cite{KriegeDavid1997TDt:} discusses equatorial platforms including some history.

\section*{Preliminary Discussion of the Roberts Linkage}

A great variety of planar linkages are known to mechanical engineers.
Some of these linkages feature a tracer point whose trajectory
describes an approximate straight line.
The Roberts linkage is one such mechanism.
Reference \cite{FergusonEugeneS1962Komf} attributes the linkage to
Richard Roberts of Manchester and dates the invention as prior to
1841.

\begin{figure}
  \caption{Roberts Linkage}
 \begin{subfigure}[b]{0.4\textwidth}
 \includegraphics[width=\textwidth]{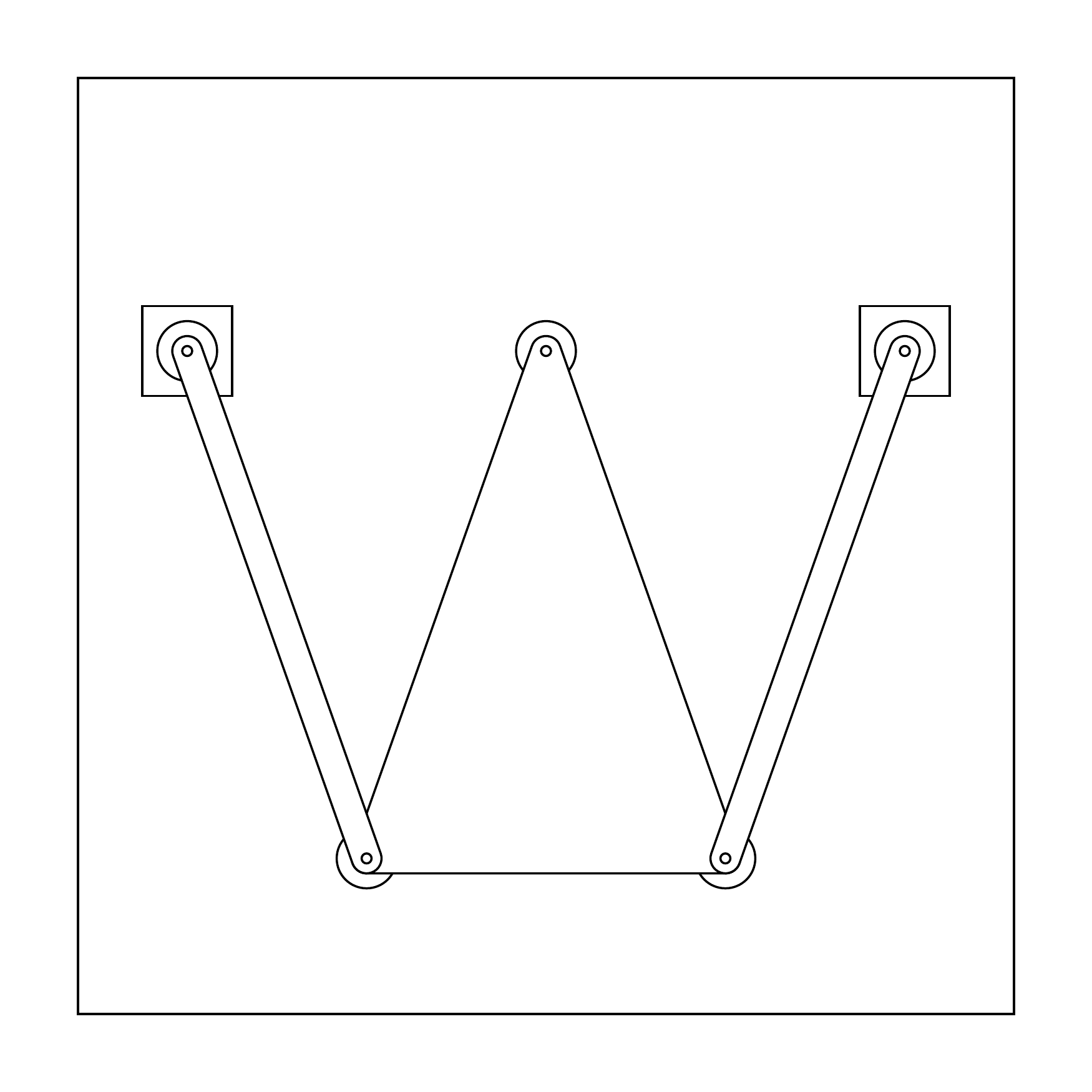}
  \caption{Centered}
  \label{fig:1}
 \end{subfigure}
 \begin{subfigure}[b]{0.4\textwidth}
 \includegraphics[width=\textwidth]{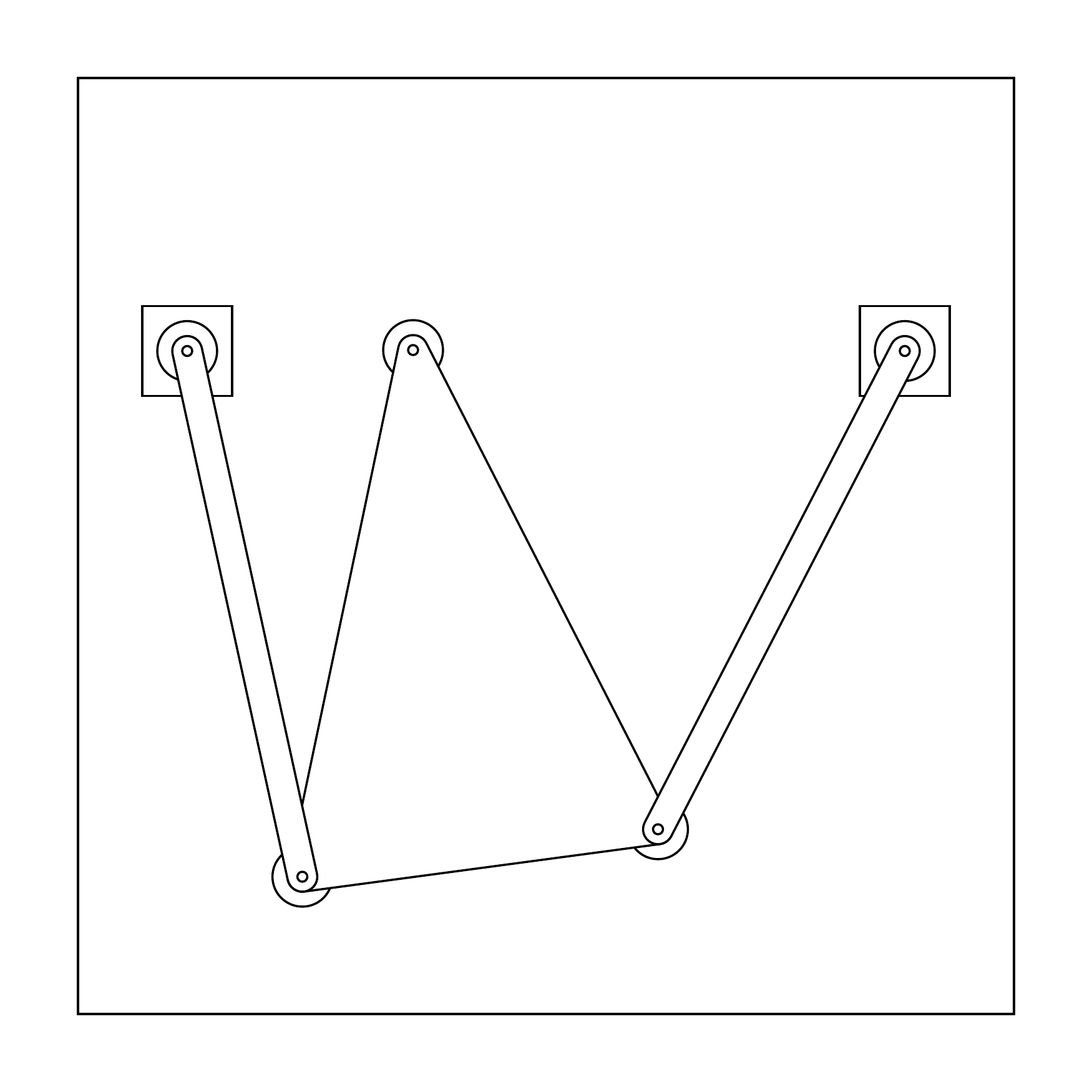}
  \caption{Inclined $7.6^\circ$}
  \label{fig:2}
 \end{subfigure}
\end{figure}

Figures \ref{fig:1} and \ref{fig:2} illustrate an example of
a Roberts linkage. Two fixed pivots are attached to two equal length struts. The other 
end of these struts are attached to two equal angle corners of an isoceles triangle
with the short side acting as a connecting link.
The equal 
long sides of the triangle have the same length as the two struts. The short
side of the triangle has length equal to half the distance between the
two fixed pivots.
In motion the vertex
of the triangle
acts as a tracing point whose trajectory 
deviates very little from a straight line.

\section*{Application of the Roberts Linkage to an Equatorial Tracking Device}
Position the plane of motion of a Roberts linkage to be same plane as the celestial equator. The normal
to this plane will be directed toward the celestial pole. In the northern hemisphere this is near Polaris.
This positioning is effectively polar alignment.

We shall call the center piece in motion the lamina.
With out violating the above positioning reqirement
we can arrange for the approximate straight line motion of the tracer point on the lamina to be horizontal.

The constrained motion of the lamina comprises both rotation and translation. If pushed at a
properly selected rate, the rotation can nullify the diurnal rotation of the earth.
This is the purpose of the present invention.
While the translational motion brings no benefit it is harmless when viewing objects
at astronomical distances.

Upon the lamina a shelf or attachment can be
installed to hold a payload telescope with altazimuth mount.

It can be further arranged that the center of gravity of
(lamina + shelf + payload) lies at a point in space which, perpendicular projected into the plane of
the lamina, lies on the tracer point.

The celestial tracking motion will result in very little vertical movement of the
center of gravity of the since the motion of the tracer point differs little from a horizontal line.

Because there is little vertical movement of the center of gravity, the gravitational
potential energy is nearly constant as the motion proceeds.

Because the gravitational potential energy is nearly constant, the force needed to
maintain the motion will be small.

\section*{Description of Existing Prototype}

\subsection{Mechanical}
\begin{center}
  \begin{figure}
    \caption{6'' $\text{F}/6$ Newtonian on the Platform}
    \includegraphics[angle=90,origin=c,scale=.12]{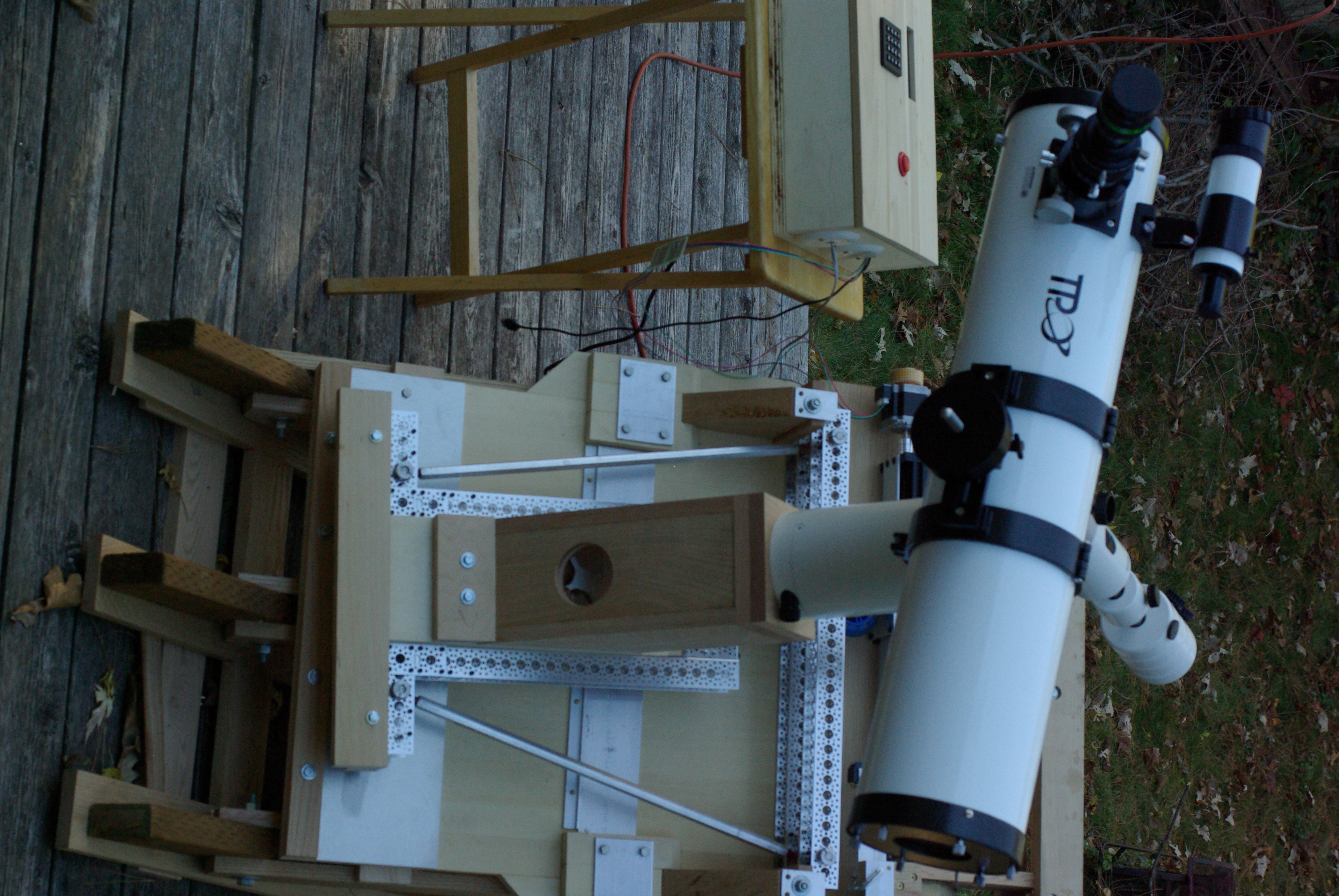}
    \label{First_JPG}
  \end{figure}
\end{center}

See Figure~\ref{First_JPG}.
A wooden frame provides a plane parallel to the celestial equator at the
observing site latitude $45^\circ$ N.

The fixed upper pivots are 21 inches apart and the connecting lower link
has distance between pivots 10.5 inches apart. The east and west side links
are 22.3125 inches long. The moving part is supported by PTFE (aka Teflon) sliders
below the lower pivots. Another slider is located on the center of the lamina
12.75 inches above the lower pivots. The surface under the sliders is Fiberglass
Reinforced Plastic (FRP).

A wheel on the moving lamina is centered and projects above the line of the upper fixed
pivots.
The height of the wheel axle above the lower two moving pivots is 25.25 inches.
This wheel is an inline skate wheel.
A linear actuator carries a plate that pushes this wheel.

A wooden square pillar stands on the lamina and attaches to a comercially available metal round pillar (Vixen part 25167).
On top of the round pillar an altazimuth mount carrying a telescope can be attached.
Care is taken with counter weights to make sure the center of gravity of the
altazimuth mount is comparativly stationary as the telescope is moved for pointing.

\subsection{Electro-Mechanical}
A Nucleo-144 STM32F767 evaluation board is the microcontroller for the project.
A 4x4 matrix keypad and 4x20 character OLED display provide input and output.
Six AA batteries power the Nucleo board independently of the 24V power supply
for the stepper driver. Opto-isolators both protect the board and provide
voltage step up from 3.3V to 5V for output signals to the stepper driver.

The stepper motor has 400 steps per revolution and the stepper driver is
set to run with 16 microsteps. The lead screw in the linear actuator
has a pitch of 8~mm.   When tracking the the microcontroller sends pulses to
the stepper driver at a freqency indicated the right hand axis of Figure \ref{fig:NonZero_NonLinearity}.

\subsection{Lessons Learned}
The full planned range of tilt of was $7.5^\circ$ east or west allowing for one hour of tracking.
This was not possible because beyond a certain point the center of gravity shifted enough
that the moving part of the apparatus began to tip over.  In practice 3~inches of east or west movement
of the push wheel seemed safe. This allows for about 30 minutes of tracking.
Note that we can space the lower sliders further apart without changing the length of
the connecting link. It is not necessary that these sliders be directly under the link.
This increased spacing will increase the range of stable motion.

Initially the PTFE sliders moved on a surface of aluminum. A vibration was visible through the
eyepiece which is likely due to slip stick friction. Switching to an FRP surface greatly reduced this.
Further plans are to polish the FRP with automobile wax.
In the next prototype the lower sliders may be
replaced with wheels.

\newpage
\appendix

\begin{center}
\LARGE{Appendices}
\end{center}

\section{Kinematical Design}
\label{Appendix:Kinematic}
Kinematical design is also called kinematic design or geometric design.
A well illustrated introduction to kinematical design is reference \cite{StrongJohn1938Piep} pp.585-590.
The guiding principle is to have no more positioning constraints on a collection of rigid
bodies than are required to fix its configuration. This involves counting degrees of
freedom of the bodies being positioned.

The Roberts linkage as employed in this invention is well suited for
kinematical design. The lamina, or any rigid body, considered as an object in three
dimesional space has six degrees of freedom. To fix its position in space
six constraints are required.

Three contact points against a plane ensure that any motion of the lamina is indeed planar.
In the existing prototype these contact points are low friction sliders and the plane is
parallel the celestial equator on a solidly constructed frame.

The east and west side links provide two more constraints. This leaves one degree of freedom
which is the tilting motion of the lamina.  The final degree of freedom can be fixed by contact with
an actuator.

\section{Solution of Configuration}
\label{Appendix:Solution}
\begin{figure}[!htb]
  \caption{Four Bar Linkage (Some say Three Bar)}
  \includegraphics[width=\textwidth]{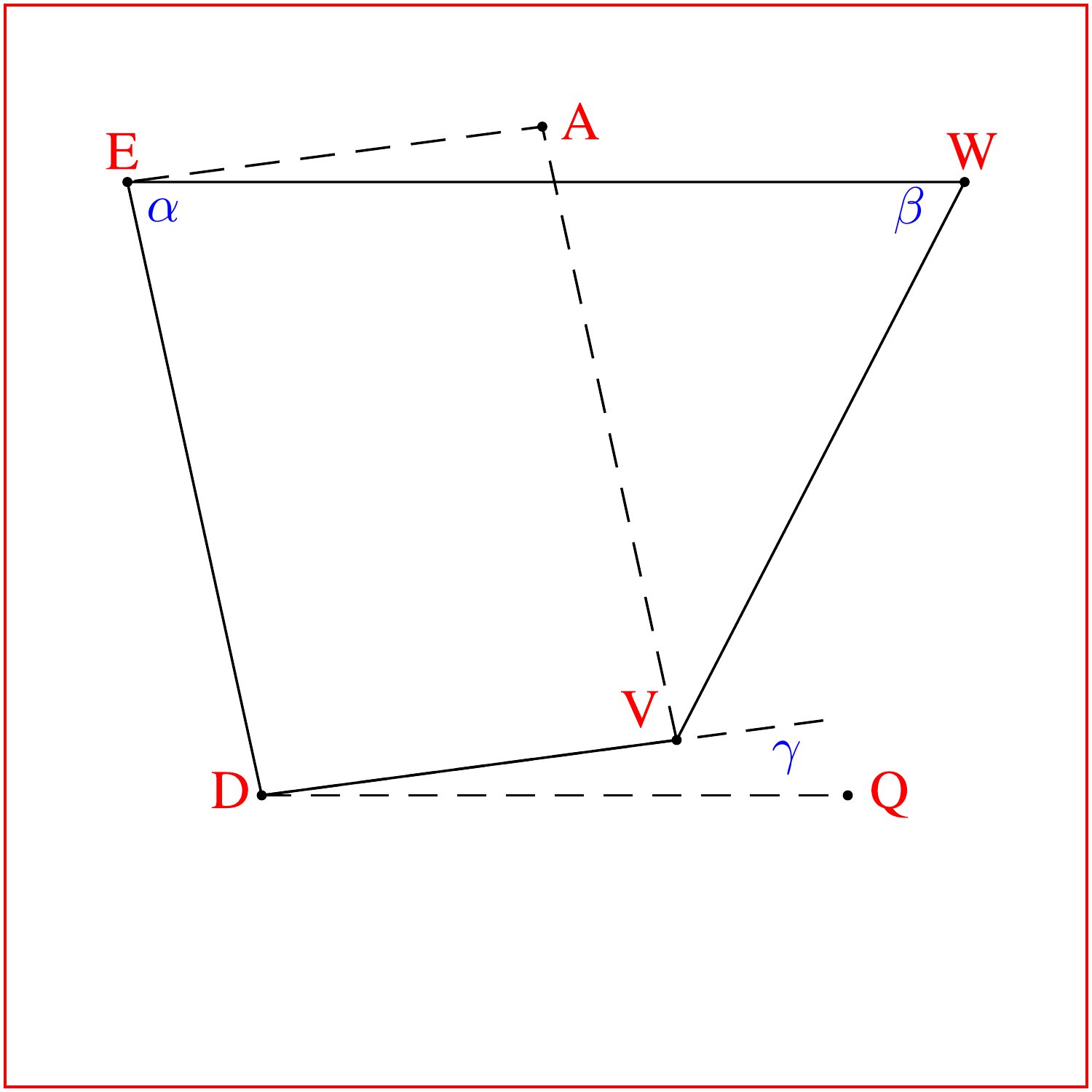}
  \label{fig:7}
\end{figure}

This solution is a variation of the trigonometric construction
shown in reference~\cite{PaulB.Burton1979Kado} p.20.
Refer to Figure \ref{fig:7}. $E$ and $W$ are fixed pivots. $D$ and $V$ are moving pivots whose configuration
is constrained by east link $ED$, west link $WV$, and connecting link $DV$.
Point $Q$ is constructed at any distance from $D$ such that $DQ$ is parallel to $EW$. Angle $\gamma = \angle VDQ$
is the input describing the tilt of the connecting link relative to $EW$.  In the present invention
we will want $\gamma$ to change at a constant rate equivilent to one revolution per sidereal day.

Given $\gamma$ we wish to determine $\alpha = \angle WED$ and $\beta = \angle EWV$.
Define fixed lengths:
\begin{align*}
   f & = \norm{EW} \\
   c & = \norm{DV} \\
   e & = \norm{ED} \\
   w & = \norm{WV}.
\end{align*}
Construct auxillary point $A$
such that $AEDV$ is a parallelogram with $EA \parallel DV$ and $ED \parallel AV$.
Note that $\angle AEW = \angle VDQ = \gamma$. Consider segment $AW$ (not drawn),
\[x = f - c \cdot \cos(\gamma) \]
\[y = c \cdot \sin(\gamma) \]
are the components of $AW$ parallel and perpendicular to $EW$.
\[d = \sqrt{ x^2 + y^2} \]
is the length of $AW$ and
\[\delta = \arctan \left( \frac{y}{x} \right) \]
is $\angle AWE$.
Most computer languages allow the evalution of $\delta$ as \texttt{atan2(y,x)}.
Applying the law of cosines to $\triangle AWV$ yields
\[ e^2 = w^2 + d^2 - 2 d w \cos( \beta + \delta ) \]
or
\[\beta = -\delta + \arccos\left( \frac{w^2 + d^2 - e^2}{2wd} \right). \]
A similar argument can show that
\[\alpha = \delta + \arccos\left( \frac{e^2 + d^2 - w^2}{2ed} \right) \]
with the same definitions of $d$ and $\delta$.
The other angles in quadrangle $EDVW$ are:
\[\angle EDV = 180^\circ - \alpha - \gamma \]
\[\angle DVW = 180^\circ - \beta + \gamma. \]
Note that the sum of all four angles, ($\alpha + \beta + \angle EDV + \angle DVW $), is $360^\circ $ as expected.

\section{Centro: The Center of Rotation}
\label{Appendix:Centro}
As discussed, in motion the lamina undergoes both translation and rotation.
One may ask, about what center does this rotation take place?
A simple construction locates this center which some textbooks call the centro.
See \emph{e.g.} reference~\cite{SmithWilliamGriswold1930Ek} p.217.

\begin{figure}
  \caption{The intersection of the red lines is the centro}
  \includegraphics[width=\textwidth]{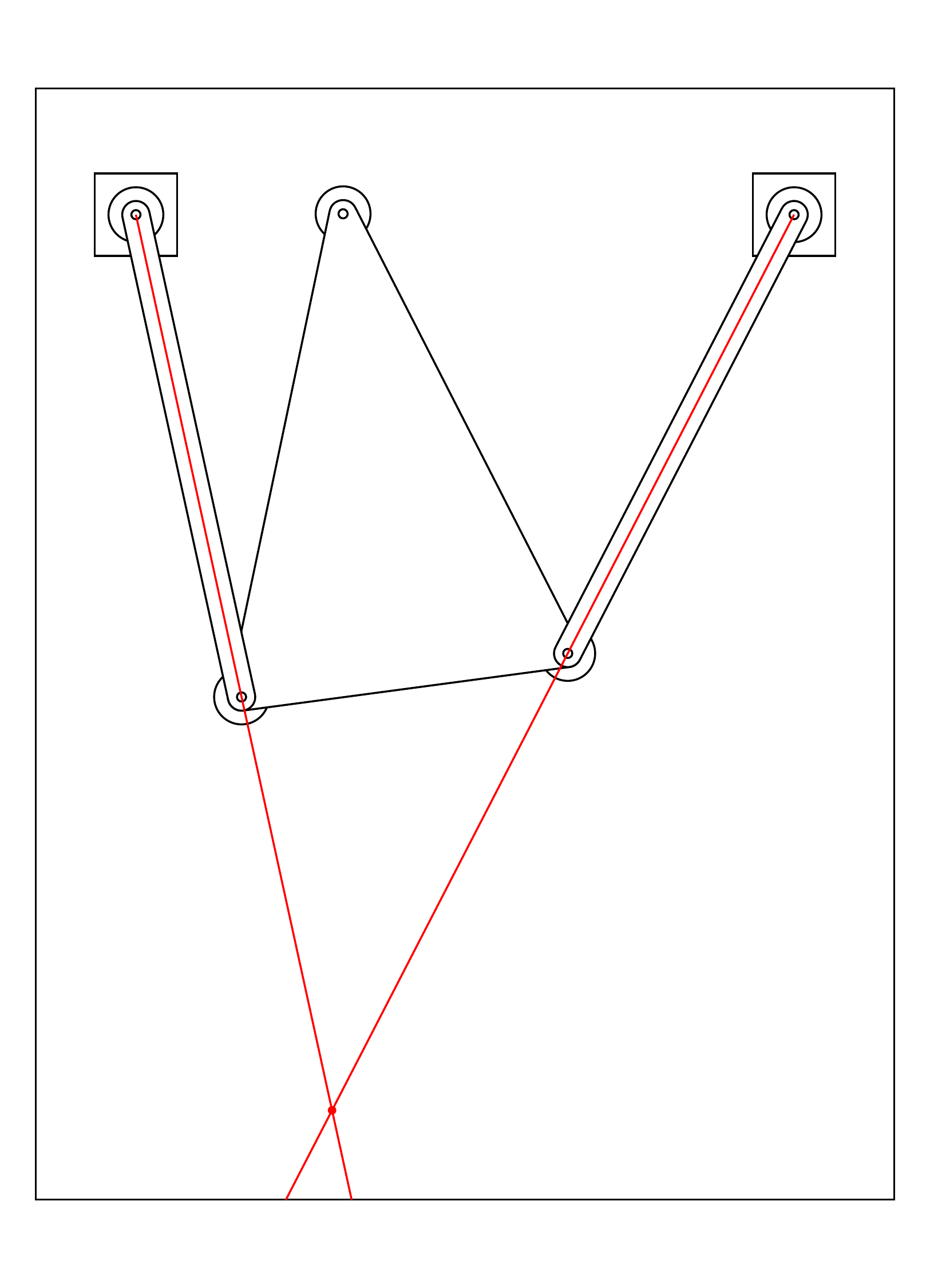}
  \label{fig:centro}
\end{figure}

Refer again to Figure \ref{fig:7}.
If we extend the line segment $ED$ representing the
east link and likewise extend line segment $WV$ representing the west link
these lines will, if not parallel, intersect at some point.
This point of intersection is the centro. This construction is
shown in Figure \ref{fig:centro}.

Suppose the triangular lamina is pushed by a horizontal actuator near the vertex
of the triangle. 
The lever arm about which rotation occurs is the displacement
from the centro to the contact point of the actuator on the lamina.
The length of this lever arm is approximately twice the altitude of
the triangle.

The comparative lever arm length of a conventional equatorial mount
is the pitch radius of the Right Ascension worm wheel.  The linkage
based platform can have a lever arm length greater by a factor of ten.
This results in greater precision which is a major advantage
of the present invention.

\section{Dealing with Nonlinearity}
\label{Appendix:Nonlinearity}
\subsection{One Way}
For the existing prototype the parameters $f$, $c$, $e$, and $w$ are:
\begin{align*}
 f & = 21.0 \text{~in} \\
 c & = 10.5 \text{~in} \\
 e & = 22.3125 \text{~in} \\
 w & = 22.3125 \text{~in}
\end{align*}
Additionally we need two parameters specifying the position of the wheel on
the lamina. To do this we need to be specific about coordinate systems used.

On the stationary frame we choose a coordinate system with origin centered on
the midpoint
of the line of the fixed pivots
with $x$ coordinate increasing to the west and $y$ coordinate
increasing upwards. In this system points $D$ and $V$ in figure \ref{fig:7} have negative $y$ coordinate.

On the moving lamina we choose a coordinate system with origin
on the midpoint
of the line of the moving pivots
with $x$ coordinate increasing to the west and $y$ coordinate
increasing upwards. 

\begin{center}
  \begin{figure}[!htb]
    \caption{The Non-Linearity is Small}
    \includegraphics[angle=0,origin=c,width=300pt]{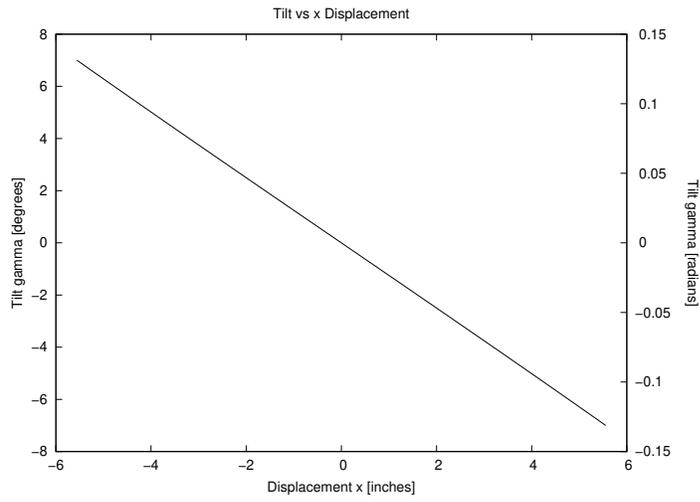}
    \label{fig:Small_NonLinearity}
  \end{figure}
\end{center}

The coordinates of the wheel axle in the moving lamina frame are:
\begin{align*}
 x & = 0.0 \text{~in} \\
 y & = 25.25 \text{~in} 
\end{align*}
The wheel so located projects above the line of the fixed pivots.
The motion of this wheel is primarily horizontal and the small vertical movement is
not perceptable.
A linear actuator pushes this wheel. The linear actuator is driven by
a lead screw rotated by a stepper motor. The calculation of Appendix \ref{Appendix:Solution}
determines the coordinates of the wheel.  Figure \ref{fig:Small_NonLinearity}
shows that the tilt $\gamma$ is nearly a linear function of the $x$ coordinate of
the wheel.

\begin{center}
  \begin{figure}[!htb]
    \caption{Plot of $\frac{dx}{d\gamma}$ reveals non-linearity}
    \includegraphics[angle=0,origin=c,width=300pt]{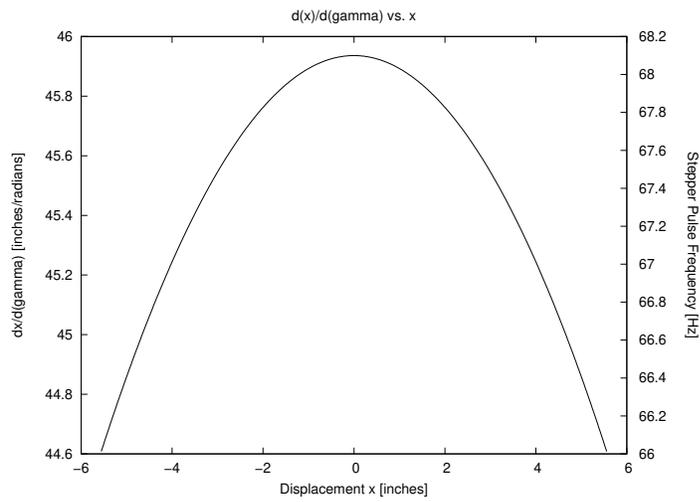}
    \label{fig:NonZero_NonLinearity}
  \end{figure}
\end{center}

We can calculate the derivatives of the push point coordinates by
treating the motion of the lamina as a small rotation about the centro.
The derivative $\frac{dx}{d\gamma}$ is plotted in Figure \ref{fig:NonZero_NonLinearity}.
This derivative is directly proportional to the rate at which the wheel must be
pushed to achieve a desired constant value of $\frac{d\gamma}{dt}$. This in turn is
directly proportional to the frequency at which pulses must be delivered to
the stepper driver.
For the existing prototype this frequency is indicated on the right hand axis of Figure \ref{fig:NonZero_NonLinearity}.

So, one way of dealing with non-linearity is to keep track of the $x$ coordinate
and send pulses to the stepper driver at a calculatable frequency dependent on $x$.

\subsection{Another Way}
The existing apparatus uses a different way of handling non-linearity.
In practice we do not specify a frequency for pulses sent to the stepper
driver. We specify a time interval. The microcontroller has various
timers which run independently of the program execution and can
generate interrupt events. These interrupts provide pulses to the stepper driver.
The data we desire is the number of timer counts between pulses.
We keep track of position of the actuator as variable \texttt{X}, the integral number of
microsteps from the center. The relationship between coordinate $x$ and \texttt{X} is
\[ x = \texttt{X} \times \frac{\texttt{THREAD\_PITCH}}{\texttt{MICROSTEPS\_PER\_REVOLUTION}} . \]

The algorithm is this.
\begin{itemize}
\item
  Program a function that provides $x$ given gamma.
\item
  Numerically invert this to have a function that provides gamma given $x$.
\item
  Use a Chebyshev representation to achieve a fast function providing gamma given $x$.
\item
  Calculate two values gamma(\texttt{X+1}) and gamma(\texttt{X}) using the Chebyshev representation.
  These numbers are angles in radians.
\item
  Divide these two numbers by $2\pi$ and multiply by the number of seconds in
  a sidereal day.
\item
  Multiply these two numbers by the frequency of the timer.
\item
  Round these two numbers to the nearest integer.
\item
  Finally subtract these numbers to obtain the number of timer ticks that must elapse between
  pulses to the stepper driver.
\end{itemize}
Rounding before subtracting avoids cumulative round off error.

An alternative setup for pushing the lamina is to have an actuator
act between a point on the lamina and a point on the supporting frame.
In this setup we can write a function that provides the distance
between these two points given gamma.
The same algorithm can
be used with this function as a starting point.

\bibliography{dans_references}

\end{document}